# Top-down predictions influence binocular rivalry through beta band rhythms: a qEEG-based investigation

Azadeh Mojdehfarahbakhsh[1], Saba Nouri[1,2], Abolfazl Alipour[3], Mohammad Nami[1,4,5,6]*

[1]Neuroscience Laboratory, NSL (Brain, Cognition and Behavior), Department of Neuroscience, School of Advanced Medical Sciences and Technologies, Shiraz University of Medical Sciences, Shiraz, Iran
[2]Students' Research Committee, Shiraz University of Medical Sciences, Shiraz, Iran
[3]Department of Psychology and Neuroscience, Indiana University, Bloomington, Indiana, USA
[4]Department of Neuroscience, School of Advanced Medical Sciences and Technologies, Shiraz University of Medical Sciences, Shiraz, Iran
[5]DANA Brain Health Institute, Iranian Neuroscience Society-Fars Branch, Shiraz, Iran
[6]Academy of Health, Senses Cultural Foundation, Sacramento, USA

*Corresponding author*
*Mohammad Nami. Department of Neuroscience, School of Advanced Medical Sciences and Technologies, Shiraz University of Medical Sciences, Shiraz, Iran. torabinami@sums.ac.ir

## Abstract

Predictive coding theory suggests that conscious perception results from the interaction between top-down and bottom-up signals in the brain. However, the electrophysiological signatures of top-down predictions are not clear yet. Here, we cued subjects to expect a certain perceptual state in a binocular rivalry task and quantitatively analyzed their EEG signals during the cue period and prior to binocular rivalry task. We found that when predictions can successfully influence the perception of the rivalrous stimuli, the power of beta band rhythms increases in primary visual cortices and beta band phase lag in a frontal-visual homologous channel pair was notably diminished. Building upon earlier works, our findings suggest that beta rhythm is potentially considered as a signature of top-down communication in the brain.

## Introduction

The effect of the brain's internal expectations on conscious perception is an important topic in neuroscience. Recent findings suggested that conscious perception results from the interaction between both top-down and bottom-up signals in the brain (1) as both top-down and bottom-up signals affect visual perception (2,3). Accordingly, finding electrophysiological signatures of top-down and bottom-up signals can pave the road to understand mechanisms of interactions between these two signal streams. In a recent study, Bastos et al. (4) used electrocorticography (ECoG) signals in monkeys and showed that bottom-up signals are relayed through theta (~4 Hz) and gamma band rhythms (~ 60-80 Hz) while top-down feedbacks use beta band rhythms for communication (~14 Hz). Additionally, Van Kerkoele et al. (5) used multisite linear electrodes and looked at the flow of feedback (top-down) and feedforward (bottom-up) signals between V1 and V4 and showed that feedforward signals are sent through gamma band (30-90 Hz) while feedback signals were propagated in the alpha-beta band (5-15 Hz). Moreover, they showed that the feedforward signals originate from superficial neocortical layers whereas feedback signals are generated in deeper layers of the neocortex (see (6,7) for a brief review of these reports).

Inspired by these findings, several studies looked at the electrophysiological signatures of top-down and bottom-up signals in humans using electroencephalography (EEG) or magnetoencephalography (MEG) measurements. Michalareas et al. (8) recorded MEG signals while subjects were performing a visual task and used Granger causality to see which frequencies are utilized for feedforward and feedback communication in visual cortex. In line with monkey studies, their results suggested that feedforward signals use gamma band and feedback signals use alpha-beta band rhythms for communication. Piantoni et al. (9) showed that reduction of beta rhythm's power correlates with perceptual reversals in a binocular rivalry task. In other words, upon observing an ambiguous stimulus that can be perceived as two different percepts, beta rhythms support the stability of one perception over another. Authors argued that when beta power goes down, long-range communication between different regions of visual cortex is reduced and a perceptual switch happens. Additionally, Keil et al. (10) showed that the power of beta band rhythms in temporal lobe sensors of MEG increases before the illusory perception of sound-induced flash illusions in subjects. This can be interpreted as a signature of global top-down effect of beta rhythms.

Meanwhile, the role of beta band rhythms in top-down 'predictions' is not a well-studied topic. Denison et al. developed a behavioral task using binocular rivalry to study the effects of predictions on visual perception of stimuli. Intriguingly, they found that presenting a stream of rotating gratings before binocular rivalry can bias the subjects to perceive the grating with an orientation that is predicted by the motion direction of the pre-rivalry grating stream (11). However, they did not record the electrophysiological signatures of this prediction process. Interestingly, Arnal et al. used MEG signals and showed that top-down predictions in an audiovisual task were relayed by beta rhythms ((12), also see (13)). Therefore, it is not known whether the same beta rhythms mediate top-down predictions in Denison et al.'s binocular rivalry task.

We hypothesized that top-down visual predictions should manifest their strength through beta band rhythms. We used a binocular rivalry task and measured the changes in beta band power and phase lag as indicators of top-down influence on lower cortical areas. Our results suggested that upon successful prediction, beta band power increase in primary sensory areas and phase lag in this frequency reduces in a frontal-visual sensor pair.

# Methods

## Participants

11 male subjects entered the study and EEG data from 4 of the subjects were used after behavioral screening (see inclusion criteria below). Mean age of the selected participants was 49±1 years old.

The experimental procedure was approved by the Ethics Committee of Shiraz University of Medical Sciences and all of the participants gave informed consent before entering the study.

## Binocular rivalry task

We used Denison et al.'s methodology for the rivalry task and modified it to be usable with a virtual reality headset. Briefly, we incorporated each block of the visual stimuli into a video clip and created a randomly ordered playlist of the video clips for each subject. Subsequently, we used a smartphone and a virtual reality headset to present the stimuli for the subjects (figure 1). Each block of the stimuli was a binocular rivalry task comprised of two parts: 1) pre-rivalry stream and 2) rivalry frame.

To build up perceptual expectations, we used a pre-rivalry stream in which a series of 15 frames each composed of two black and white circular grating patches with a spatial frequency of 3 cycles per degree (cpd) were projected to both left and right eyes of the subjects. A gray border and a fixation sign were placed on both left and right eye stimuli to facilitate visual convergence [Denison et al]

During the pre-rivalry stream, gratings were rotating coherently for 45 degrees every 300 milliseconds either counterclockwise or clockwise. This movement direction in the pre-rivalry stream was counterbalanced.

In order to see the effects of the perceptual expectation on binocular rivalry, we presented subjects with a rivalrous static frame for 4 seconds at the end of each pre-rivalry stream where one of the eyes received a grating that was congruent with the expectation from the pre-rivalry sequence frames and the other eye received the opposite grating that was incongruent with the movement direction of pre-rivalry sequence frames. The eye which was receiving congruent figure (left or right) and the orientation of the final figure (45°, 135 °, 90 °, 180°) were counterbalanced.

According to these counterbalancing arrangements (2 different eyes, 2 rotation direction, 4 final orientations), there were 16 trials per experiment. We repeated the experiment for 5 times which resulted in 80 trials per participant in total.

Subjects verbally reported their first stable and dominant perception in the rivalry frame and their responses were checked to see if they are consistent with predicted motion direction of pre-rivalry frames or not. EEG data from rivalry epochs were discarded due to the noise introduced to the data through verbal reports and their irrelevance for research question of this study. We named those perceptions that are congruent with the motion direction of gratings as 'successful prediction trials' and the rest of the trials were called 'unsuccessful prediction trials' throughout the experiment.

## Participant inclusion/exclusion criteria

To control for the eye dominance (the fact that some participants perceive the rivalrous stimulus that is projected on a particular eye regardless of other conditions in most of the

trials(11), we adopted Denison et al.'s exclusion criteria and excluded subjects who perceived rivalrous gratings that were projected on one of their eyes for more than 85% of the time. This reduced the number of participants in this study from 11 to 4.

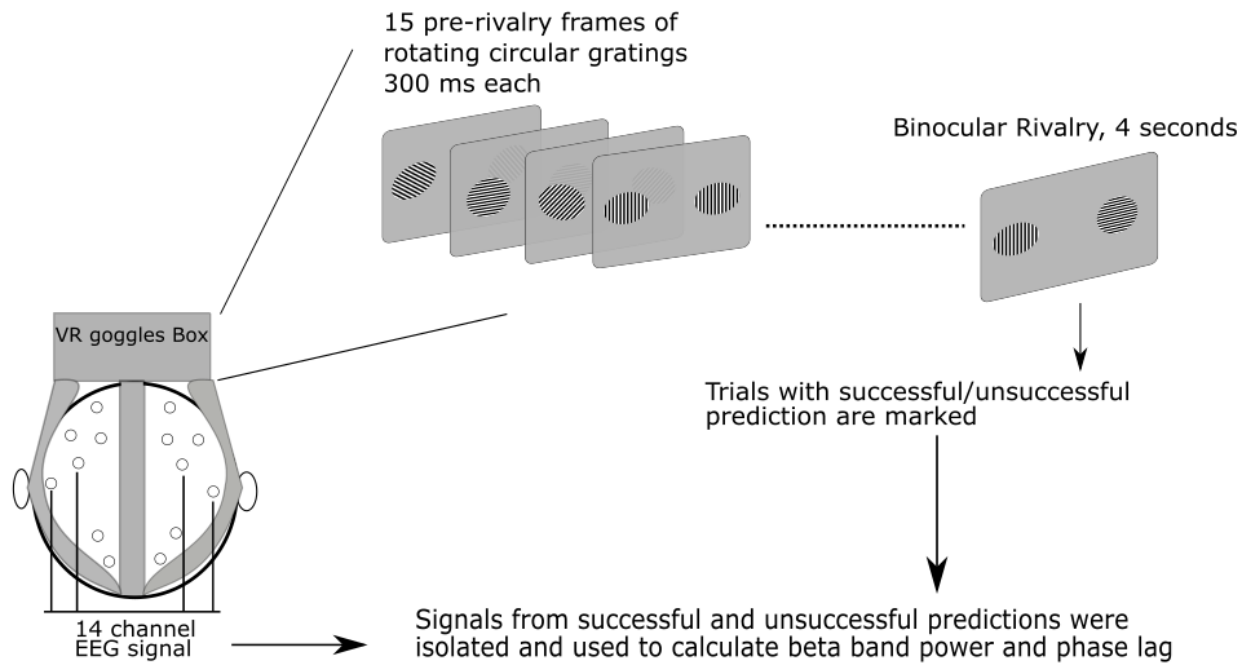

**Figure 1.** Schematics of the experimental procedure. Subjects viewed 15 rotating circular gratings (pre-rivalry stream) and then they were given a binocular rivalry task. Subjects reported their first dominant and stable perception in the binocular rivalry. Trials in which perception of the subjects matched the motion direction of pre-rivalry gratings were marked as successful trials and the rest of the trials were marked as unsuccessful. EEG signals during pre-rivalry stream in both successful and unsuccessful trials were extracted, cleaned, and used to calculate beta power and phase lag.

### EEG recording

To record brain activity using EEG signals during task performance with minimal interference with subject ability to use the virtual reality headset, we used Epoc+ EEG headset (Emotiv, USA) and recorded EEG signals throughout the task performance. Emotiv Epoc+ EEG headset is shown to have an acceptable signal to noise ratio and signal quality compared to other research-grade EEG systems (14). The EEG signals were recorded from the following channels: AF3, F3, F7, FC5, T7, P7, O1, O2, P8, T8, FC6, F4, F8, AF4 with a sampling rate of 128 Hz. The starting timepoint for each recording session was annotated by experimenter and used as a reference point for data analysis.

Data analysis:

EEG data was converted from EDF to ASCII format and imported to EEGlab (15) running on Matlab 2018b. EEG signals were assigned into four different groups based on the rivalry results (i.e. perception was [in]congruent with rotation direction) and orientation of the rivalrous stimuli in the final frame (whether it was 45° VS. 135 ° or 90 ° VS. 180°). Then, to isolate brain dynamics during pre-rivalry stream, EEG signals during binocular rivalry epochs (4 seconds) were removed.

Subsequently, large amplitude artifacts were manually removed from the data. Afterwards, to remove eye and muscle artifacts from the data, we ran an ICA analysis- using runica.m from EEGLab- on the EEG dataset and noisy components were visually identified (through their spectral features and localization patterns) and they were removed subsequently. After removal of noisy components, ICA was run again to ensure the removal of all noisy components in the pruned dataset. Noisy components were still identifiable after the first round of ICA analysis in two of the datasets and they were removed in a second round of ICA artifact removal.

Subsequently, to find spectral signatures of top-down modulation in visual cortex, we looked at the changes in the power of beta band rhythms (15-30 Hz) in electrodes close to primary visual areas (i.e. O1, O2, P7, P8) when perception of binocular rivalry is matched with pre-rivalry stream motion direction. We pooled the percentage changes in the power of beta band rhythms from all four electrodes (O1, O2, P7, P8) and used a paired *t*-test (ttest.m on Matlab) for statistical comparison of changes in the beta band activity over primary visual areas.

Additionally, in order to observe changes in interaction between higher and primary cortical areas, we exported the ICA pruned data from EEGLab and imported it to Neuroguide QEEG software [also explained in (16)] and calculated the phase lag between channels in each hemisphere at beta band. Then, we used a paired *t*-test (ttest.m on Matlab) to see if the phase lag between intra-hemispheric electrode pairs changes during successful predictions. Specifically, we looked at the changes in beta band phase lag between frontal channels and the two channels on primary sensory areas in each hemisphere (i.e. phase lag between (AF3-F3-F7-FC5) and (P7-O1) for left hemisphere and between (AF4-F4-F8-FC6) and (O2-P8) for the right hemisphere). This generated phase lag values in each successful and unsuccessful prediction trials. Values from unsuccessful trials were subtracted from successful trials to obtain phase lag changes between them. Consequently, phase lag changes were used for statistical comparison and to find significant changes.

## Results

### Increased beta activity during successful prediction

Beta band activity (15-30Hz) showed a significant increase in primary visual areas during successful prediction trials (paired *t*-test, $p<0.05$). Figure 2 shows scalp map of subject No. 1 in two conditions of unsuccessful vs. successful prediction.

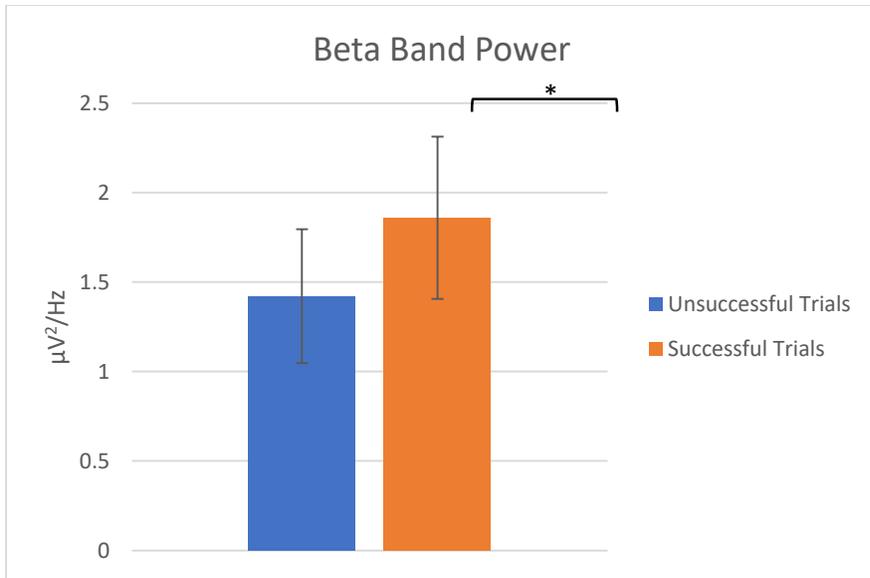

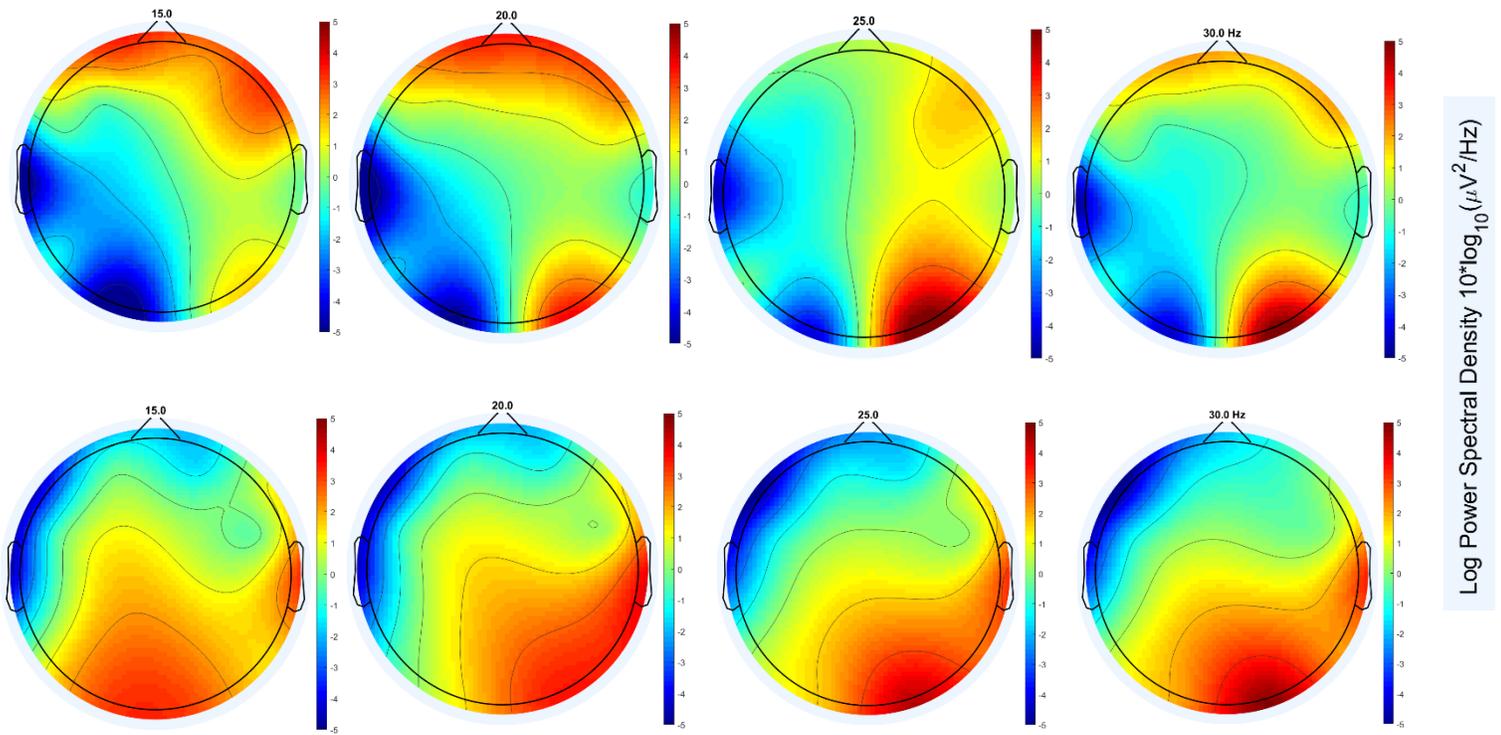

**Figure 2.** Increased beta band power during successful prediction trials (data from subject No. 1). A. Beta band power in successful trials increased compared to unsuccessful trials (pooled across P7, O1, O2, P8). B. Scalp map from subject one in successful (bottom) and unsuccessful (top) trials (pooled across all trials that ended with rivalry between 45° and 135° grating patches).

## Reduced phase lag between F3-O1 electrode pair during successful prediction

We found that the O1-F3 electrode pair showed a significant increase of phase lag during successful prediction trials (from -1° in unsuccessful trials to +16° in successful trials, paired $t$-test, $p<0.05$).

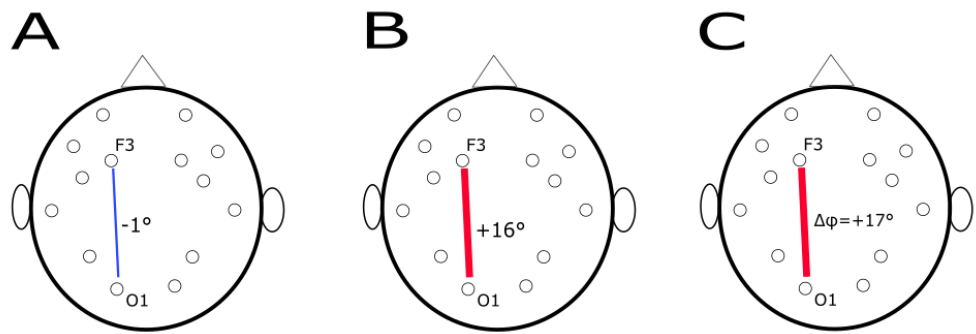

**Figure 3.** The Phase lag between F3 and O1 shows a significant increase when predictions successfully affect the results of binocular rivalry. A and B are mean phase lag in unsuccessful and successful prediction trials, respectively. C. Difference between the successful and failed prediction trials.

## Discussion and conclusion

Our results suggest that the power of beta band rhythms increases in primary sensory areas and the phase lag between O1-F3 pair increases when top-down predictions can effectively influence visual perception in a binocular rivalry task. This corroborates the proposal that feedback connections modulate perception through beta band rhythms by sustaining the current state of the brain (17) or even reactivation of neuronal ensembles at lower levels (18).

Meanwhile, Volberg et al. (19) found that theta band rhythms are responsible for top-down signaling in a contour grouping task and suggested that beta rhythms might be acting as a local mechanism for grouping rather than being a frequency band for long range top-down modulation. Moreover, Romei et al. (20) used transcranial magnetic stimulation (TMS) to modulate the activity of the parietal cortex and they found that stimulation at the beta band favors local information processing rather than long-range top-down modulation. In a similar vein, it was shown that in a bistable stimulus that could be either perceived as the local motion of dots or a global Gestalt motion, perception of the global Gestalt is associated with a reduction of parietal beta rhythms in EEG recordings.

Therefore, the idea that top-down modulation is relayed through beta band rhythms is still unknown and a unifying mechanism for the role of beta rhythms in the brain is still needed. Spitzer and Haegens (18) divided models of beta band formation into two different categories: the first category of models suggest that beta band rhythms are generated in basal ganglia and then propagate in the cortex through thalamus whereas the second category of models suggest that beta rhythms are generated within the local cortical circuits. One of the recent cortical beta models by Sherman et al. (21) suggested that beta rhythms are generated through interaction between subthreshold synaptic currents at apical and distal dendrites of pyramidal neurons in the neocortex and showed that this model is able to cortical replicate beta bursts (18,21). Sherman et al. concluded that their model can account for the generation of beta rhythms in different brain areas and different species (monkeys, humans, and mice) and beta rhythms support top-down cortical communication.

Therefore, we believe that our results have direct consequences for this framework by showing that the beta rhythms are involved in top-down predictions of sensory stimuli in the brain. Moreover, top-down and bottom-up interactions may happen through cross-frequency coupling between beta and gamma rhythms as suggested by Craig et al. (22). Similarly, it was suggested that theta-gamma cross-frequency coupling can subserve interhemispheric communication between parietal channels (23).

In conclusion, beta rhythms seem to be signatures of top-down predictive signals in the brain. Our results suggested that when predictions can successfully influence perceptions, beta rhythms have a more powerful presence in primary sensory areas. Further research using more advanced brain imaging technologies and causal interaction analysis in source space during predictions can shed more light on the role of beta rhythms in top-down predictive processes.